\documentclass[twocolumn]{svjour3}          
\smartqed  
\usepackage{graphicx,hyperref,amssymb,amsmath,ulem,soul,xcolor}
\newcommand{\MWhe}{MWh$_\mathrm{e}$}
\newcommand{\MWht}{MWh$_\mathrm{t}$}
\begin{document}

\title{Potential Early Markets for Fusion Energy\thanks{The authors' affiliation does not
imply endorsement of this work by ARPA-E, DOE, or the U.S. Government.}
}


\author{Malcolm C. Handley* \and Daniel Slesinski$^\dagger$ \and Scott C. Hsu}

\authorrunning{M. C. Handley, D. Slesinski, and S. C. Hsu} 

\institute{
            Advanced Research Projects Agency--Energy (ARPA-E)\\
            U.S. Department of Energy\\
            Washington, DC 20585, USA\\
            \email{scott.hsu@hq.doe.gov}\\
            *Malcolm C. Handley is now unaffiliated\\
            $^\dagger$Daniel Slesinski is now with Johns Hopkins University}


\maketitle

\begin{abstract}
We identify potential early markets for fusion energy and their
projected cost targets,
based on analysis and synthesis of many relevant, recent
studies and reports. 
Because private fusion companies aspire to start commercial deployment
before 2040, we examine cost requirements for
fusion-generated electricity, process heat, and hydrogen production based on today's market prices but
with various adjustments relating to possible
scenarios in 2035, such as
``business-as-usual,'' high renewables penetration, and carbon pricing 
up to 100~\$/tCO$_2$.  
Key findings are that fusion developers should consider focusing
initially on high-priced
global electricity markets and including integrated thermal storage in order
to maximize revenue and compete in markets with high renewables penetration. Process heat
and hydrogen production will be tough early markets for fusion, but may open
up to fusion as markets evolve and if fusion's levelized cost of electricity falls below 50~\$/\MWhe. Finally, we discuss potential ways for a fusion 
plant to increase revenue
via cogeneration (e.g., desalination, direct air capture, or
district heating) and to lower capital costs (e.g., by minimizing construction times and interest or by retrofitting coal plants).  
\keywords{Fusion energy \and Energy markets}
\end{abstract}

\section{Introduction}
\label{sec:intro}

It is widely assumed that when fusion energy gain is demonstrated, humanity 
will be on the cusp of an age of economical, abundant, and carbon-free 
energy. Indeed, there are reasons to believe that if fusion power is achieved
and allowed to mature, low costs might follow.
Assuming that technical feasibility is demonstrated, several things must 
happen for fusion to reach maturity: regulations must not stifle it; the 
public must accept it; and governments and investors must support it. Most 
importantly, fusion must find early markets that are large and 
profitable enough to continue to develop. Contributing to deep decarbonization may then follow.
In this paper, we identify fusion's potential early markets and their cost targets. 

Many private fusion companies aspire to start commercial deployment before 2040, and recently the National Academies of Sciences,
Engineering, and Medicine recommended that the ``Department
of Energy and the private sector should produce net electricity
in a fusion pilot plant in the United States in the 2035--2040 timeframe''~\cite[p.~ES-2]{NAS21}.
Therefore, we examine cost requirements for
fusion-generated electricity, process heat, and hydrogen production based on today's market prices but
with various adjustments relating to possible
scenarios in 2035, such as ``business as usual,''
high renewables penetration, and  
carbon pricing up to 100~\$/tCO$_2$. We do not consider carbon pricing $>100$~\$/tCO$_2$
because direct-air carbon capture and sequestration may cost as low as 100~\$/tCO$_2$~\cite[ch.~5]{NAS19}. We also consider increased market 
size and decreased cost of producing hydrogen relative
to today.
For power generation, we consider only grid-scale markets.
Although we examine possible cost requirements and conditions in 2035, this should not be construed
as a prediction of when fusion will be deployed.

Our study is conservative.  We ignore markets that require large changes in 
the energy sector or 
existing infrastructure because these add to the many commercialization risks that fusion 
already faces. We take a somewhat unforgiving view of the competitive 
situation. We assume that other energy technologies will continue seeing
large cost 
reductions, and we do not explicitly take credit for any
potential benefits 
of fusion over other energy technologies.
We caution the reader to keep these caveats in mind. If and when fusion is 
proven at scale with reasonable costs (even if initially higher than the 
competition), it may be a disruptive energy technology that could 
eventually fundamentally alter markets and the way humans use energy. 
Fusion may well deliver on the promise of clean, abundant, safe, affordable 
energy and replace many other energy sources. Its long-term 
markets are potentially enormous.

Researched and written in support of the Tech-to-Market (T2M) component
of the ARPA-E fusion portfolio \cite{nehl19,alpha,bethe,gamow}, this paper is intended to be an informational
resource both for ARPA-E fusion performers and for potential fusion 
investors, to help them focus on initial markets that can nurture fusion 
through its early deployments. It is also for the broader fusion 
research 
community, to whom we hope to clarify the characteristics that fusion may 
need, at least initially, in order to succeed in the marketplace.   
While fusion is the main focus of this report, much of the analysis is
relevant to advanced nuclear fission. 

Finally, regulation and licensing will clearly impact fusion development, cost, and time to market.
Fusion regulation is an important subject that is
outside the scope of this paper.  We refer the reader to recent white papers \cite{KLGates18,roma20,FIA20,pillsbury20} and 
presentations \cite{NRC20} on the subject.

The paper is organized as follows.  The next section, ``Key Findings,'' is an executive summary
of the paper, followed by sections that detail the key findings on ``Electricity,'' ``Process Heat,'' ``Hydrogen and Its Derivatives,'' and ``Economic Boosts.''  The paper closes with a ``Summary.''

\section{Key Findings}
\label{sec:findings}
  
Energy markets are fiercely competitive and have low margins. However, there are many specific markets with higher energy prices that can serve as beachheads for fusion.  Key findings of this paper are summarized in this section. Details, analysis, and
references supporting these findings are given in the following sections.

Dollar amounts quoted in this paper, unless otherwise noted, are generally from 2015--2020. The relative difference among quoted
values due to inflation is at most 10\%, which is smaller than other uncertainties. Therefore,
unless otherwise noted, all quoted dollars can
be considered to be in 2020 dollars. We use today's prices rather than
projected prices in 2035, which have large uncertainties, vary across sources, and are the subject of academic research in their own right.

We use levelized cost of electricity (LCOE) to compare the cost of different energy sources because it is the most complete and widely accepted measure, including construction cost, the cost of capital during construction, the lifetime of a plant, and the cost of operating and maintaining the plant.

\paragraph{Electricity} is a promising initial market and can potentially support fusion with LCOE as high as $\sim 100$~\$/\MWhe\ in some global regions. 
Electricity is a commodity, which means that specific offtake partners do not need to be found for each project. It can
be traded, unlike process heat, though not globally, and hence is subject to significant regional price variation. This variation is driven by different fuel costs and, increasingly, different levels of renewables penetration and carbon pricing, and might make some markets initially more viable/profitable for fusion. We find that there could potentially be a very large market for fusion power plants with an LCOE of 50~\$/\MWhe.

\paragraph{Integrated thermal storage} will help fusion compete on
power grids dominated by solar and wind, which bid at 0~\$/\MWhe\ for much of the day.
Thus, fusion must be able to deliver a competitive LCOE even while selling electricity for as little as 12 hours per day. Because of the high capital and fixed costs and low variable costs of a fusion power plant, a capacity factor of 0.5 would set a high LCOE\@. Onsite thermal storage allows the fusion core to run continuously while 
selling power only when the grid price is high, enabling lower overall LCOE\@. Thermal storage appears to be a low-cost addition to a fusion 
power plant. 
In these situations, direct generation of electricity from fusion (i.e., without heat as an intermediary) may be less economically beneficial than previously expected.

\paragraph{Process heat} may be a difficult early market for fusion.
It is a hyper-localized market to which fusion power plants must be tailored.  The fusion power plant would likely need to be sited next to facilities that utilize the heat. These constraints result in
slower and riskier deployment of new heat sources. Most of the market will initially be inaccessible to fusion because of high-temperature requirements, the fulfillment of which by fusion will require additional materials research and development.  In addition, because many heat-generation
processes run off fuel that is produced as a byproduct of the process,
cost targets are likely to be below reach for early deployments of fusion.

\paragraph{Hydrogen production} may also be a difficult early market for fusion.
Hydrogen prices vary less than electricity prices but perhaps by enough to create some potentially viable early markets. For example, if fusion can get to 78~\$/\MWhe\ or below, then it may become feasible to use fusion to produce hydrogen as a partial substitute for natural-gas heating 
of buildings.

\paragraph{Additional revenue sources} can improve the economics of fusion power plants. 
Cogeneration or use of waste heat from a fusion plant to power direct-air carbon-capture equipment, desalination equipment, or district-heating networks can significantly improve the overall economics. Modeling suggests that the effective cost of energy can be lowered by as much as 35\% in some cases.

\paragraph{Retrofitting/repowering coal plants} might save 30\% of the capital cost of fusion
power plants by reusing many of the balance-of-plant components. 

\paragraph{Carbon pricing} will help fusion relative to
the fossil-fuel competition. Although there are many markets where fusion could be competitive without carbon pricing, the latter expands the markets for fusion. One fifth of the world’s greenhouse-gas emissions are already covered by a carbon price, ranging from insignificant to over 100~\$/tCO$_2$ in Scandinavia \cite[p.~15]{world_bank}.
A price of $\ge 20$~\$/tCO$_2$ could allow fusion to be competitive with fossil fuels almost everywhere if fusion costs $\le 50$~\$/\MWhe. The impact of carbon pricing on the hydrogen market is less significant.

\section{Electricity}
\label{sec:electricity}

Grid electricity is a promising initial market for fusion because electricity is a commodity and has significant price variations, including some regions with much higher prices. Furthermore, a fusion power plant will not require customization for a particular customer. Electricity is traded enough that a power plant is not dependent on one nearby customer, but not traded so much that a few ultra-low-cost players dominate the global market. Fusion may be able to penetrate most markets if it can eventually reach costs of $\le 50$~\$/\MWhe.
Beyond the factor of cost, areas best suited for the first fusion power plants may be those with higher population densities and lower land
availability/suitability for large-scale renewables generation.

\subsection{Early Markets}

Table~\ref{tab:electricity} shows a sampling of wholesale electricity prices
from around the world.
Because the first fusion power plants will likely be more expensive than most incumbent forms of power, specific high-priced markets may be best suited for early fusion deployments. 

\begin{table}[tb]
\caption{Regional wholesale electricity prices \cite[Tab: Global wholesale prices]{SE}.}
\label{tab:electricity}
\begin{tabular}{lp{.33\linewidth}cc}
\hline\noalign{\smallskip}
Market & Benchmark & Price & Market \\
 & & (\$/\MWhe) & (GW)\\
\noalign{\smallskip}\hline
Singapore & Average wholesale price & 110 & 6\\
Japan & Average wholesale price & 92 & 108\\
U.S. & Northern CA wholesale price & 61 & 38\\
Poland & Baseload wholesale price & 60 & 17\\
Italy & Baseload wholesale price & 59 & 33\\
U.K. & Baseload wholesale price & 55 & 35\\
Slovenia & Baseload wholesale price & 55 & 2\\
Portugal & Baseload wholesale price & 54 & 5\\
Spain & Baseload wholesale price & 53 & 27\\
Estonia & Baseload wholesale price & 51 & 1\\
U.S. & Average wholesale price & 39 & 445\\
World & Power from gas, coal, and nuclear (2035) & -- & 2283\\
\noalign{\smallskip}\hline
\end{tabular}
\end{table}

Singapore could be one such location, where in 2018 the average wholesale electricity price was 110~\$/\MWhe,
with 95\% coming from natural gas.
In addition, being one of the most densely populated areas of the world, Singapore will need an energy source that can be located near or within the city, while delivering large amounts of power with little land usage.

With a wholesale electricity price of 92~\$/\MWhe\ and with limited land
for renewable energy, Japan may be another favorable country for early fusion deployment. In addition, Japan 
has a high level of technological capability to facilitate rapid fusion deployment.

In the U.S., northern California has a wholesale electricity price of 61~\$/\MWhe, which is considerably higher than the rest of the country.  New England is
a candidate early market as well because it is not ideal for onshore wind nor solar additions due to a lack of sunny days and large swaths of open land. In addition, the northeast as a whole has the highest population density in the country, with many large, power-hungry cities.

Much of Europe could become available if fusion costs $\le 50$~\$/\MWhe.

\subsection{Capacity Payments}

Using storage and transmission to provide reliable power from variable renewables can add considerable costs. As a result, many U.S. electricity markets offer capacity payments to firm (i.e., not variable) power plants to encourage their construction.
These payments are 75~\$/kW$_\mathrm{e}$ per year in New England and have recently been 100~\$/kW$_\mathrm{e}$ per year on the PJM grid
(\url{https://www.pjm.com/about-pjm}) \cite[p.~33]{ingersoll20}. 
At these levels, such
payments can make a significant contribution to the economics of a fusion power plant, reducing the LCOE by 10 and 13~\$/\MWhe, respectively \cite[Tab: Capacity Payments]{SE}.
Thus, choosing a location with high and predictable capacity payments could significantly improve the economics of a fusion power plant.


\subsection{High Renewables Penetration}

Grids with high renewables penetration present a challenge for fusion
because the marginal cost of highly intermittent renewables production is zero.
The result is that whenever renewables are producing power, they meet the entire grid
demand and are likely the lowest-cost power source available, but there are long periods when renewables produce little or no power.

\subsubsection{Integrated Storage}

Integrated thermal storage can help make fusion competitive on grids with significant renewable penetration. 
The traditional response to variable, low-price power from renewables is to ramp down other power plants (e.g., natural-gas plants) during times of high renewables production. However, 
unlike natural-gas plants, a fusion power plant will have high capital and fixed costs relative to its fuel costs, meaning that relatively little money is saved when a fusion plant idles.

Consider a hypothetical future fusion plant that faces a similar situation as the example of
a fission power plant in southern California with a capacity factor of only 0.67 in
a high-renewables scenario
\cite[p.~42]{ingersoll20}. Table~\ref{tab:thermal_storage} illustrates 
the economics of such a scenario. 
The “Default plant” column describes a plant with 0.90
capacity factor and LCOE of 33~\$/MWh. As the capacity factor 
drops to 0.67 and then 0.5 because of variable, low-cost 
renewables, the plant LCOE rises by 48\% and then 97\%, respectively,
relative to the baseline \cite[Tab: Storage]{SE}.

\begin{table*}[tb]
\caption{Impact of thermal storage \cite[Tab: Storage]{SE},
where the ``Default plant'' is based on the ''average'' case of
\cite{woodruff20}.}
\label{tab:thermal_storage}
\begin{tabular}{p{0.12\linewidth}p{0.145\linewidth}lcccc}
\hline\noalign{\smallskip}
Type & Item & Units & Default plant & High renewables & Higher renewables & With storage\\
\noalign{\smallskip}\hline
Input & Rated power & MW & 400 & 400 & 400 & 400\\
Input & Capacity factor & \% uptime & 90 & 67 & 50 & 90\\
Fusion capex & -- & \$/MW & 2,366,538 & 2,366,538 & 2,366,538 & 2,866,538\\
Storage capex & -- & \$/MWh & -- & -- & -- & 35,000\\
Ten-hour storage capex & -- & \$/MW & -- & -- & -- & 350,000\\
Capex total & -- & \$/MW & 2,366,538 & 2,366,538 & 2,366,538 & 3,216,538\\
Opex & Fixed O\&M & \$/MWyr & 8,520 & 8,520 & 8,520 & 11,580\\
Opex & Variable O\&M & \$/MWh & 0.95 & 0.95 & 0.95 & 1.29\\
Opex & Fuel (assume free) & \$/MWh & 0 & 0 & 0 & 0\\
Output & LCOE & \$/MWh & 33.15 & 49.01 & 65.36 & 45.06\\
Output & LCOE relative to default & \% & 100 & 148 & 197 & 136\\
\noalign{\smallskip}\hline
\end{tabular}
\end{table*}

However, if a hypothetical fusion plant can divert its output to storage, then the fusion core can continue to run at high capacity factor instead of idling during periods of high production by variable renewables. When the grid price for electricity rises, the plant can sell stored energy as well as the energy that it continues to generate. For example, a fusion core might power a plant that delivers, for limited periods, a peak power that is double that of the fusion core alone, with the difference coming from thermal storage. 
A National Renewable Energy Laboratory (NREL) cost model for concentrated solar power with molten salts \cite{turchi13} suggests that ten hours of storage for a 400-MW$_\mathrm{e}$ plant would cost \$140M or 350,000~\$/MW\@.
The capital cost will also be higher due to the higher peak
capacity of the turbine and associated equipment. These
higher costs are included in the
last column of Table~\ref{tab:thermal_storage}, which
shows that the same plant fitted with thermal energy storage performs better than the default plant in the face of high renewables penetration. Sufficient storage to cover 10--12 hours of production could allow for all-day operation of the fusion core,
although the size and other requirements of such thermal storage must be carefully considered \cite{bubelis19}. However, if the operations and maintenance cost of a fusion plant becomes a big
enough fraction of the LCOE (e.g., from costly periodic replacement of the first
wall due to neutron damage), then idling might
become more economical than adding thermal storage.
We did not explore this, but the reader
may explore the sensitivities using \cite[Tab: Storage]{SE}.

\begin{figure}[!b]
\includegraphics[width=3.2truein]{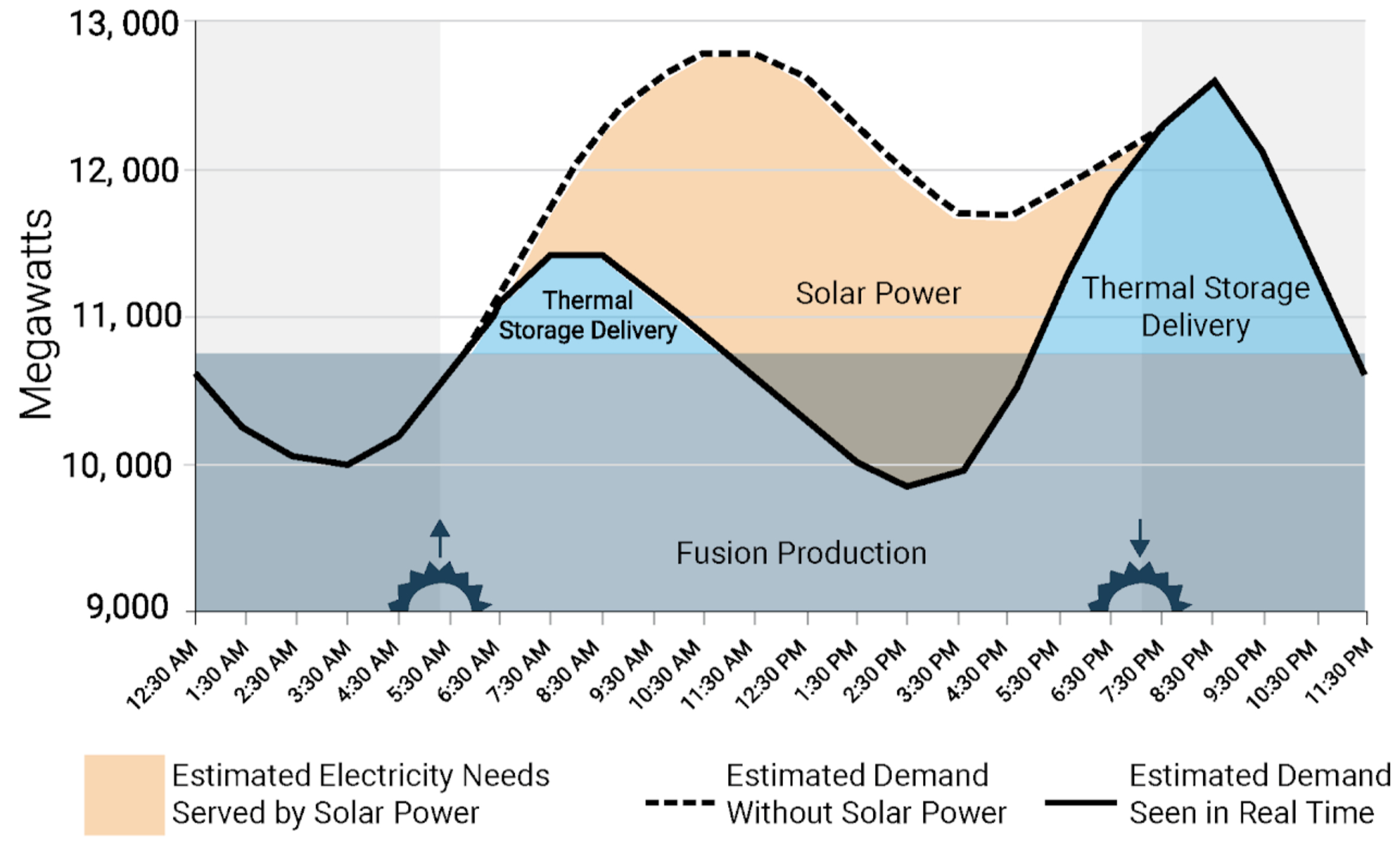}
\caption{``Duck curve'' with fusion and thermal storage.
Figure is adapted from ISO New England.}
\label{fig:duck_curve} 
\end{figure}

Figure~\ref{fig:duck_curve} shows how this might work in a case with plentiful solar power. In the early morning (when fusion generation is higher than demand) and in the afternoon (when fusion intersects with solar generation), the excess energy
is stored. When neither solar nor fusion alone can meet the total demand, the latter can be met by fusion plus thermal storage.

Firm energy sources like fusion need
less storage than variable renewables to integrate well with
the grid.  This is because firm sources only need enough storage
to allow for variations in demand while variable sources
must also accommodate variations in supply, which
can occur on a much longer time scale, e.g., seasonally.  Thus,
storage connected to firm power can tolerate higher capital
cost because it is utilized more often, providing better
amortization.  See \cite[p.~37]{ingersoll20} for an
economic analysis of storage coupled with fission power plants.

\subsubsection{Unviability of Using Curtailed Fusion Energy
to Produce Hydrogen}

Using curtailed fusion electricity to generate hydrogen when renewables are over-producing does not appear to be economically viable.
During these times, 
the excess power from renewables is unused and available for purchase. Any consumer would do better to buy renewable power instead of more expensive fusion power.

\subsection{Accessing Larger Markets}

The ideal outcome for fusion in the longer term is to have a lower LCOE than
the competition, as has been the case recently with natural gas costing less than coal. In this subsection, we examine market prices as a function of the prices of natural gas and coal, as well as the penetration of renewables. We find that for fusion to be competitive in most markets, its LCOE should be in the range 40--50~\$/\MWhe.

For fusion to be competitive in many large markets, it may need to reach a similar LCOE to that of natural gas. While the LCOE of natural-gas generation will vary based on the fuel cost, as examined below, it is useful to consider present LCOE estimates of natural
gas, which have a range of 44--68~\$/\MWhe\ \cite[p.~2]{lazard19}, as a benchmark.   For new natural gas coming into service in 2025, the EIA reports LCOE of 38~\$/\MWhe\ for natural gas combined cycle (NGCC) and 67~\$/\MWhe\ for combustion turbine \cite[p.~7]{EIA20}. This shows that an aggressive benchmark for the cost of fusion would be about 40~\$/\MWhe.

While the above costs are a useful reference, it is important to further note that the price for natural gas will vary based on location, with the fuel being a significant portion of the total cost. A study
from Google, which estimates the cost of integrating intermittent and dispatchable electricity sources on a grid \cite{platt17}, finds that a fusion-like (i.e., hypothetical firm, low-carbon energy source) power plant would undercut all other energy sources in the U.S. market, including existing plants, if it could produce electricity at 17~\$/\MWhe even when natural gas costs as little as 2.5~\$/MMBTU. This would potentially open up a 308-GW market to fusion plants as fast as they can be built. The Google study also finds that with a less-aggressive LCOE of 28~\$/\MWhe, a fusion-like system could replace existing power plants as they are retired. This would create a demand of 12 GW of new fusion plants per year.  See \cite[Tab: Replacing existing assets]{SE} for calculations of these numbers.

\begin{figure}[tb]
\includegraphics[width=3.2truein]{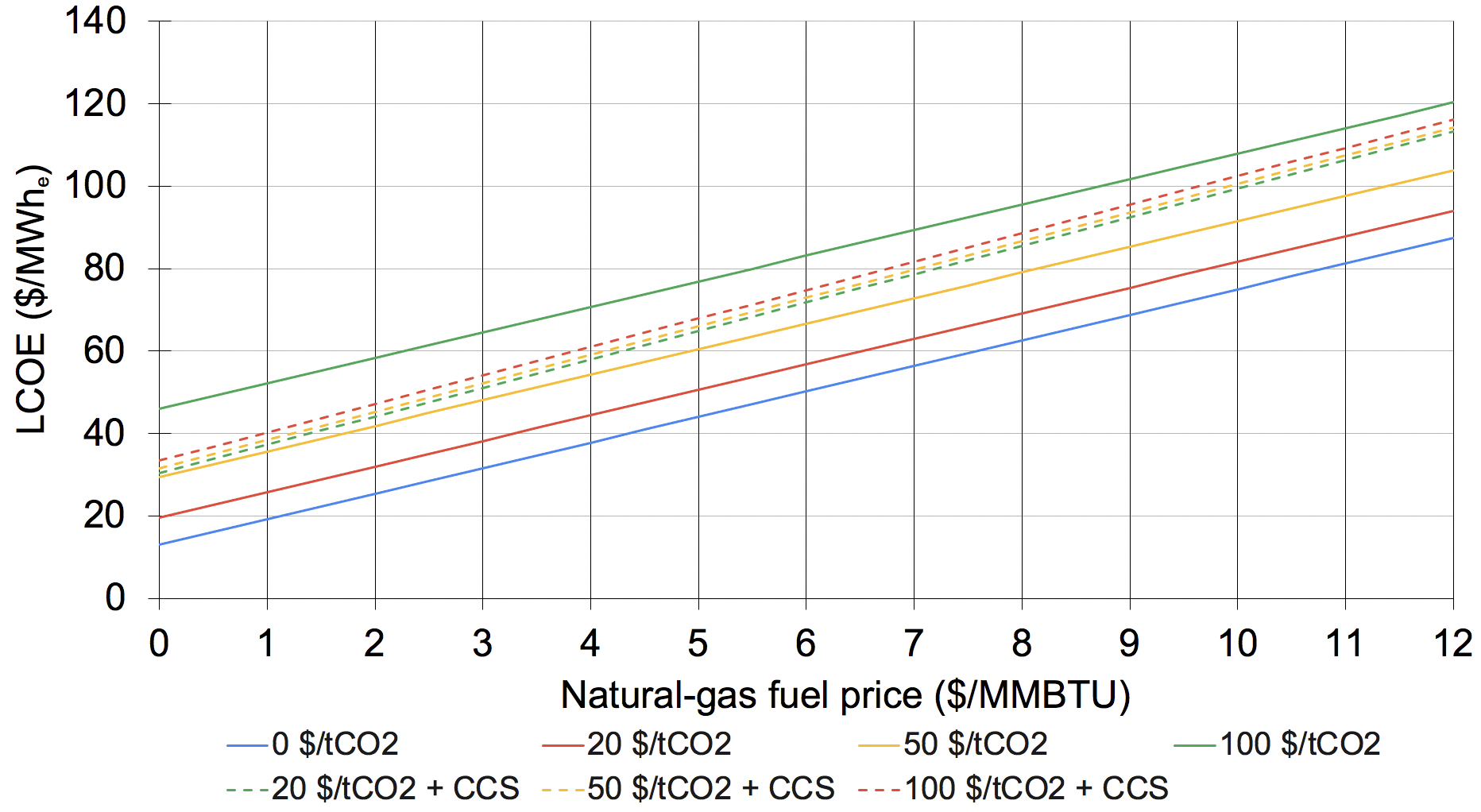}
\caption{LCOE for NGCC generation for various carbon and fuel prices \cite[Tab: LCOE for NGCC]{SE}.}
\label{fig2}       
\end{figure}

Regional variation in the price of electricity is driven by variation in 
fuel costs, carbon price, and renewable potential. Using cost assumptions 
from the U.S. market, we examine the price of competing power sources as a
function of these inputs \cite{platt17}.
Figure~\ref{fig2} relates the price of natural-gas generation to the price of fuel, the carbon price, and whether carbon capture and sequestration (CCS) is used. Natural-gas prices in 2019 were about 2.5~\$/MMBTU in the U.S., with the highest being 10~\$/MMBTU in Japan; only in the U.S. and Canada are prices below 4.4~\$/MMBTU \cite[p.~39]{BP20}.  Figure~\ref{fig2}
suggest that fusion at $\lesssim 50$~\$/\MWhe\ can potentially
become competitive with natural gas over a range of projected scenarios.

\begin{figure}[tb]
\includegraphics[width=3.2truein]{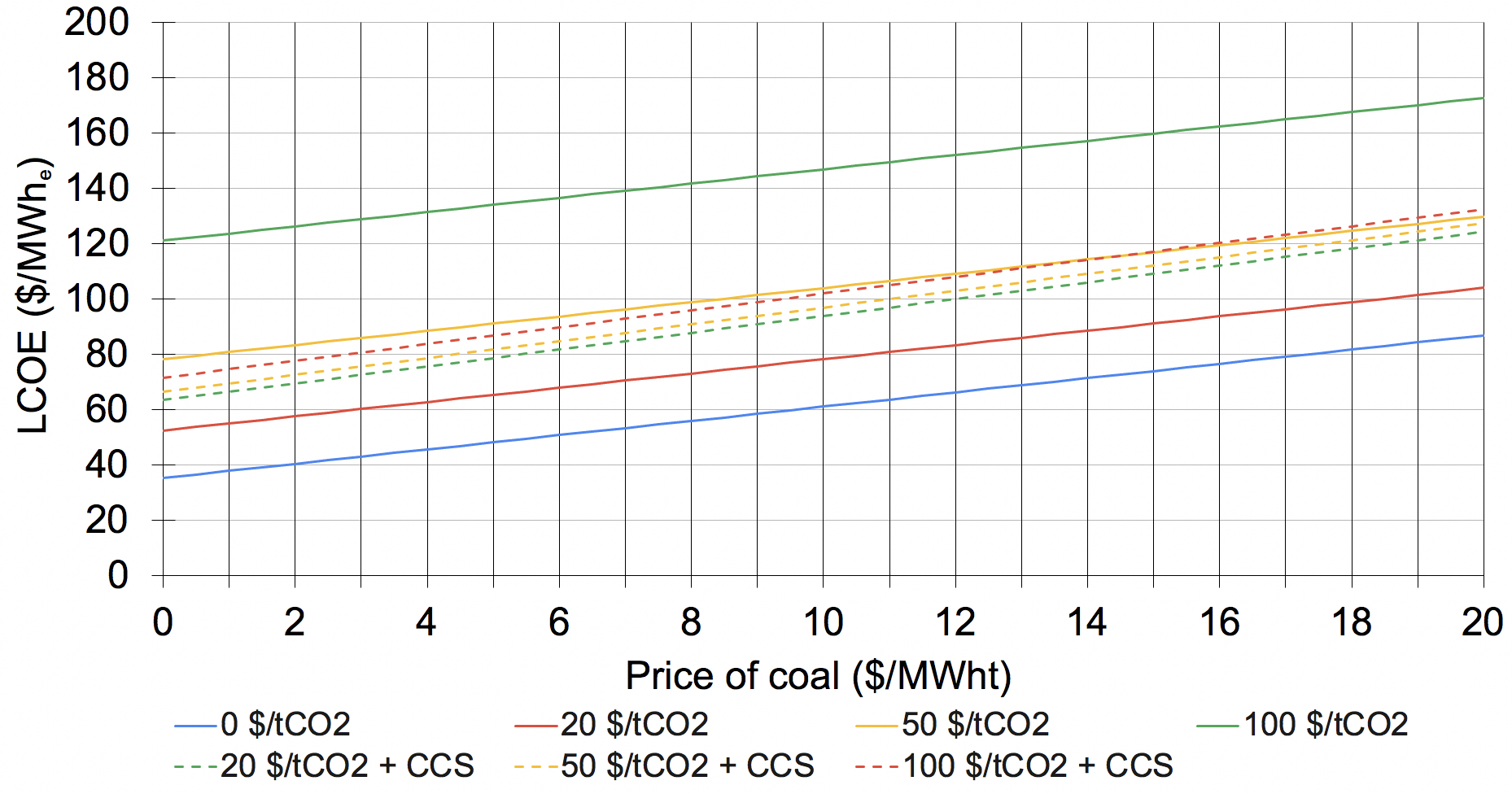}
\caption{LCOE for coal generation for various carbon and fuel prices \cite[Tab: LCOE for Coal]{SE}.}
\label{fig3}       
\end{figure}

The final firm power source we consider as a competitor to fusion is coal. In Fig.~\ref{fig3}, we relate the price of electricity from coal power plants to the price of coal, the carbon price, and whether CCS is used. Over the last 20 years, coal prices have been as low as 4~\$/\MWht\ and as high as 20~\$/\MWht\ \cite{ritchie17}, ignoring the price for coking coal. In 2018, prices ranged from 9~\$/\MWht\ in the U.S. to 11~\$/\MWht\ in Europe to 14~\$/\MWht\ in Asia.  These data suggest that
fusion at $\lesssim 60$~\$/\MWhe\ can potentially outcompete coal over
a range of projected scenarios.

\begin{figure}[tb]
\includegraphics[width=3.2truein]{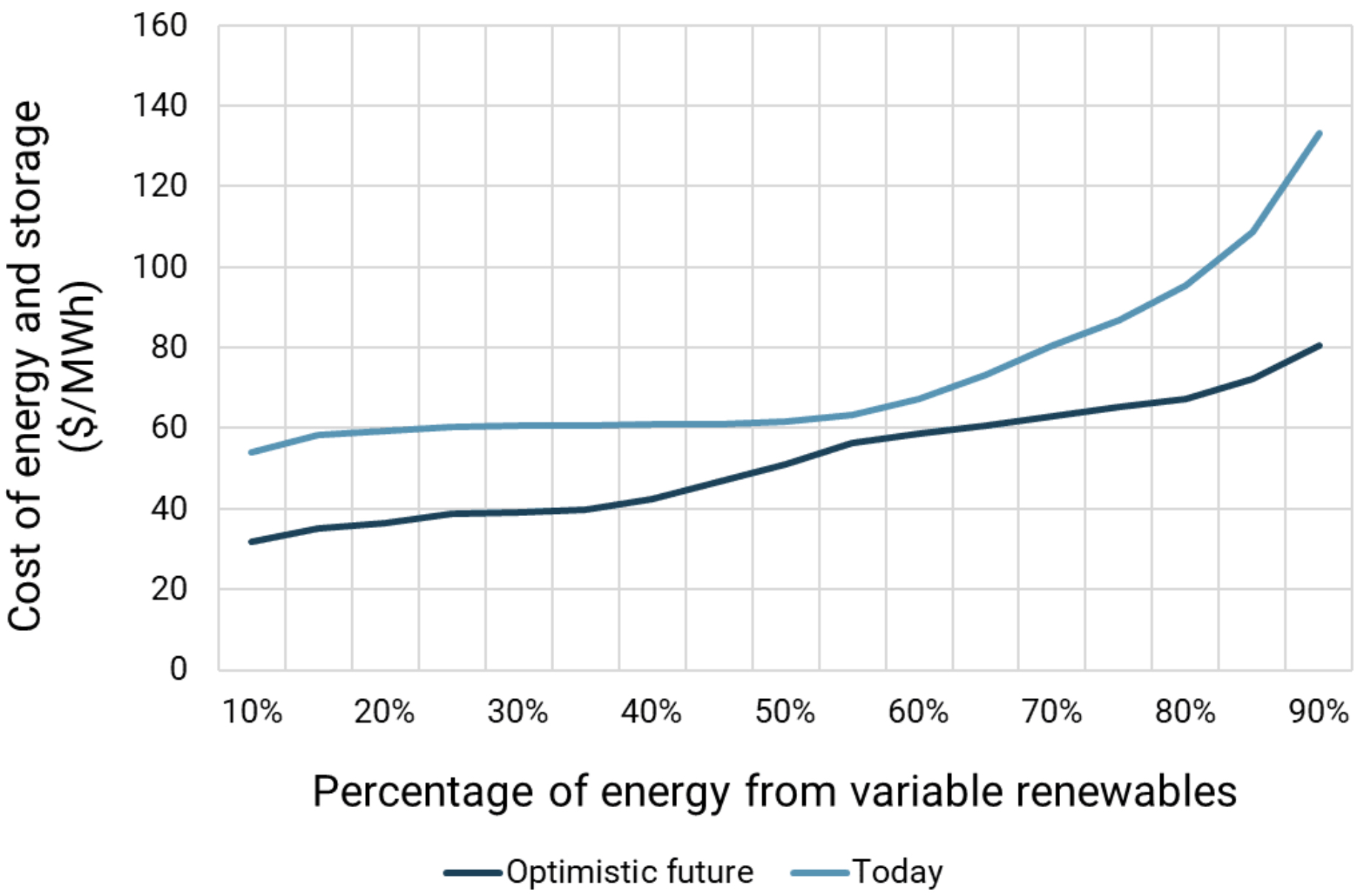}
\caption{LCOE versus renewable fraction in California \cite[Tab: SCOE for renewables]{SE}.}
\label{fig4}       
\end{figure}

Wind and solar produce very low-cost power, but the latter is not always the same as a low system-wide LCOE\@. Rising penetration of renewables leads to rising marginal costs because of the mismatch between the times and locations of supply and demand, and the cost of transporting and storing energy, as shown in Fig.~\ref{fig4} for California.
The future capital and fixed costs for solar are 0.63~\$/W and 21.66~\$/\MWhe, respectively, and 1.37~\$/W and 46.71~\$/\MWhe, respectively, for wind \cite{ritchie17}.  Recent work has shown that including a firm, low-carbon
source (like fusion) would reduce overall system cost in deep-decarbonization
scenarios for power generation \cite{sepulveda18}.

Based on the analyses of this subsection, we conclude that fusion with an LCOE of 41~\$/\MWhe\ should be able to compete with natural gas in most countries, coal in almost all countries, and renewables when they make up 35\% or more of the grid. With a carbon price of at least 20~\$/tCO$_2$, fusion at 50~\$/\MWhe\ can compete with gas in most countries other than the U.S., coal in all countries, and renewables when they are 50\% or more of the grid. If fusion can reach a low enough price to compete with natural gas, it could be an excellent complement to high levels of renewables. By 2035 and beyond, more regions will likely have high penetration of renewables, and these could be good early markets for fusion. 

\subsubsection{Modularity for Growth Markets}

Selling to countries with the largest growth in electricity demand may require providing modular equipment or using a build-own-operate (BOO) model.
Building large infrastructure projects anywhere is risky, and such projects often go over their original budget and/or schedule.
Considering energy factors such as how much clean energy a country already has, economic factors such as GDP, and institutional factors such as regulatory quality, a recent study \cite[p.~11]{ford20} finds that 85\% of demand growth will be in countries with below-median economic and institutional readiness.
Fusion will find more success in these markets, and more rapidly, if plants are small and modular, so that their output can be more easily absorbed by the grid and they require less from the receiving country in terms of finance, construction, and oversight. Failing this, different deployment models may help. For example, a fusion company that can offer a BOO service (where it builds, fuels, operates, and decommissions the plant) greatly reduces the burden on the receiving country.

\section{Process Heat}

Process heat is used in a wide range of industrial processes, including chemical production, oil and steel refining, and paper production. It is a large market,
comprising 22\% of primary energy use in the U.S. in 2014 \cite[p.~188]{MIT18}. However, it is a challenging market that will require special circumstances to be accessible to fusion
as an early market.  For example, a facility that faces unusually high energy costs, such as a remote mine, could be an excellent candidate for early fusion deployment.

One challenge is that process-heat plant designs vary widely, even within the same industry \cite[p.~2]{bloomberg19a}, and may require a custom retrofit in order to accept heat from fusion. Another challenge is that plant operators often want to wait for several successful demonstrations in their industry before they consider wide-scale adoption.  We have heard from many experts that the process-heat industry is very conservative, eschewing risks even when there is an opportunity to save money. 
Finally, consumers of industrial heat often require temperatures (e.g., $>700^\circ$C) that may be challenging for fusion (or nuclear in general) to deliver
(due to the temperature limitations of low-activation structural
materials in a neutron-irradiation environment),
or they use effectively free fuel that is a byproduct of their production process. Still, an industrial plant might choose to have a colocated fusion power plant to supply electricity for electrical process heating.

The size of the process-heat market is best described 
in \cite[Appendix~F]{MIT18}, which presents a detailed analysis of the potential for fission to provide process heat,
considering the amount of power needed by each plant, the temperature required, and whether the plant produces its own fuel as a byproduct of its operation.  However, economics, siting restrictions, and less tangible issues such as public opinion are
all ignored. Thus, the conclusion that only 19\% of the market, totaling 637~GW$_\mathrm{t}$ worldwide \cite[pp.~26--27]{MIT18}, is a fit for fission heat ignores most of the characteristics that might give fusion an advantage.
In addition, the study makes two assumptions that may not be true for fusion. First, fission can provide heat up to 850$^\circ$C \cite[p.~187]{MIT18}. In principle, fusion power plants should have the same maximum temperature range as fission plants (limited by materials properties), but if the maximum temperature is higher or lower for fusion, the achievable part of the heat market will be larger or smaller, respectively. Second, the report assumes that fission power plants can be as small as 150--300~MW$_\mathrm{t}$. Many fusion designs may need to be larger (dictated by physics) or combine several modules of smaller size at a single site in order to amortize the cost of the tritium-processing plant. The minimum unit size (with respect to power generation) of a single
fusion core is not yet known and the subject of ongoing research.

\section{Hydrogen and Its Derivatives}

Hydrogen can be a substitute for fossil fuels for a wide variety of purposes, and can be used to make other chemicals that can also substitute for fossil fuels. For example, hydrogen can be used instead of natural gas for refining oil and iron and for heating buildings, and be turned into ammonia for fertilizers and methanol for synthetic gasoline.
Today, hydrogen is almost exclusively made from fossil fuels. If zero-carbon hydrogen (``green hydrogen'') can be produced with fusion power at competitive prices, then this could be a very large market for fusion. The price of hydrogen is more consistent around the world than the price of electricity, but still varies enough to allow for some promising potential early markets for fusion.
Hydrogen production uses a combination of heat and electricity but tends to be produced in large facilities that can justify operating their own power plant. As such, the market is different enough to warrant exploring separately.

In this section, we examine methods for producing hydrogen and plausible costs for producing it with fusion power, and then look at the more significant uses for hydrogen. We discuss market sizes in megatons of hydrogen (following the lead of the literature), which can be converted to a market for fusion energy by noting that producing hydrogen with electrolysis uses about 50~kWh$_\mathrm{e}$/kgH$_2$ (or 5.6~GWyr/MtH$_2$).

\subsection{Producing Hydrogen}

Hydrogen can be produced from hydrocarbons (using them both as a fuel and a source of hydrogen) or from electrolysis of water
(using electricity to split water into hydrogen and oxygen). Without a carbon price, the 
lowest-cost way to produce hydrogen is from natural gas or coal. With a high enough carbon price and improvements in electrolysis and renewable energy, the price ceiling would be set by electrolysis powered by renewable energy.

In summary, we find several potential early markets for hydrogen produced with fusion power. Hydrogen in China and Europe costs about 2.3~\$/kgH$_2$ when made from natural gas with CCS
\cite[p.~42]{IEA19}. Green hydrogen made with fusion-powered electrolysis could be competitive if the fusion power plant’s LCOE is $\le 50$~\$/\MWhe\ (for fusion-LCOE estimates, see ``Electrolysis" subsubsection below). Without CCS, hydrogen from natural gas costs about 1.7~\$/kgH$_2$ in these regions \cite[p.~42]{IEA19}, which would require a challenging fusion LCOE of 36~\$/\MWhe. Western Europe and Japan together could blend nearly 5~MtH$_2$/year into existing natural-gas networks for heating without significant modification, with some fraction of this market becoming available if fusion costs 78~\$/\MWhe\ (Japan) or 66~\$/\MWhe\ (Europe).
However, the above assumes an absence of large-scale competition from wind and solar. Including
the latter, fusion-powered electrolysis may require an LCOE of $\le 32$~\$/\MWhe\ to 
produce hydrogen competitively.  Details are discussed below.

\subsubsection{Natural Gas and Coal}

\begin{figure}[tb]
\includegraphics[width=3.2truein]{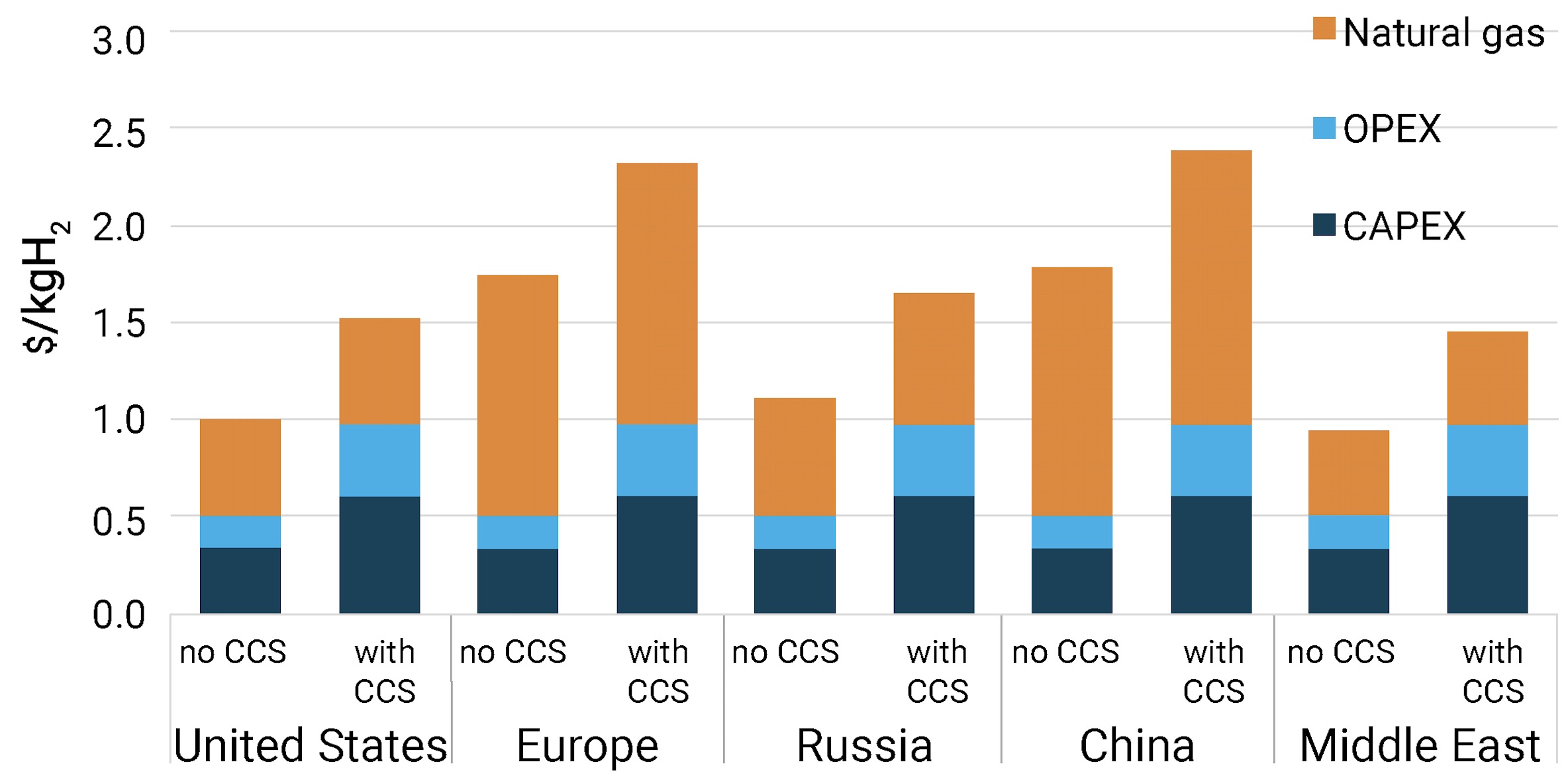}
\caption{Regional variation of hydrogen prices \cite[p.~42]{IEA19}.}
\label{fig5}       
\end{figure}

Today, almost all hydrogen is made from natural gas by pressurizing it and heating it with steam to break the hydrocarbons, or from coal by heating coal and water to form ``syngas,'' and then separating out the CO$_2$ \cite[p.~42]{IEA19}. These two processes are called steam methane reforming and coal gasification, respectively. 
The cost of these processes varies based on the local costs of gas and coal, but is in the range of 0.71--2.29~\$/kgH$_2$
\cite[p.~11]{bloomberg20}. Figure~\ref{fig5} shows the regional variation of these prices.

\subsubsection{Electrolysis}

Using electricity to split water to produce hydrogen
is energy-intensive.  However, with a low-enough cost of energy and high-enough capacity factor, hydrogen produced by electrolysis can become competitive with hydrogen produced from natural gas or coal. Using the assumptions described below, fusion LCOE would need to be 13--50~\$/\MWhe\ to be competitive with fossil-fuel production of hydrogen.

\begin{figure}[tb]
\includegraphics[width=3.2truein]{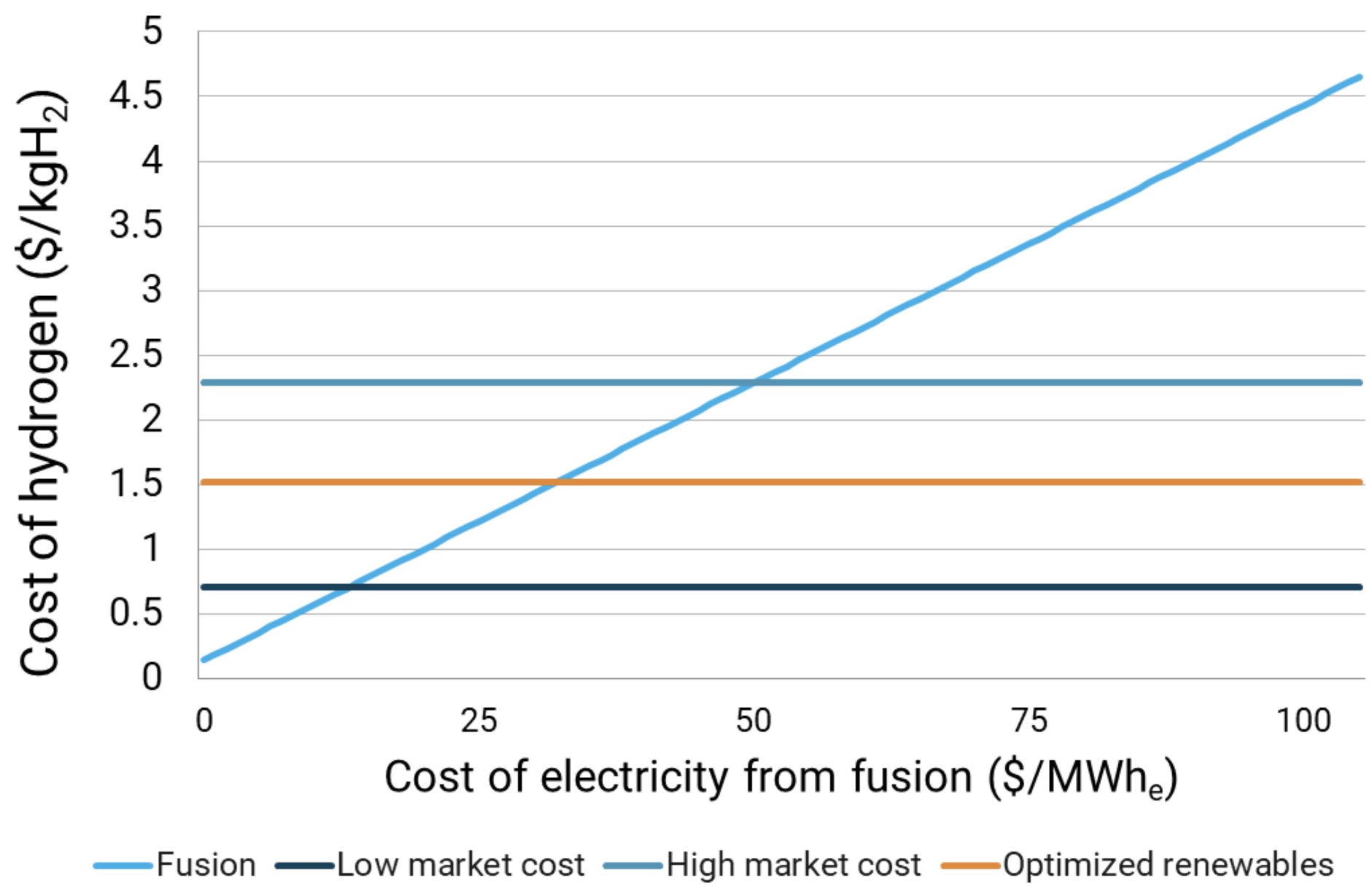}
\caption{Cost of producing hydrogen from fusion-powered electrolysis \cite[Tab: Production, Column: Fusion, low-temp]{SH}.}
\label{fig6}       
\end{figure}

To determine this, we adjusted the model in  
\cite{bloomberg19b} to describe electrolysis powered by a fusion power plant. 
Because our calculations do not exactly match the prices given in \cite{bloomberg19b}, as detailed in 
\cite[Tab:  Production, Row: Correct value for line above]{SH}, all of the costs presented here are from our model, even for scenarios modeled in \cite{bloomberg19b}.
Electrolysis is presently very expensive (6--7~\$/kgH$_2$ in North America and Europe 
\cite[p.~28, Fig.~18]{bloomberg19b}), but several significant improvements over the baseline are described (see
Table~\ref{tab:electrolysis_improvements}) and included in our model, as they would all likely be implemented when integrating hydrogen and fusion plants.
Figure~\ref{fig6} shows the low and high market costs for fossil-fuel production of hydrogen, corresponding to fusion LCOE of 13 and 50~\$/\MWhe, respectively.

\begin{table}[tb]
\caption{Possible improvements for cost of electrolysis \cite{bloomberg19b}; corresponding page numbers from \cite{bloomberg19b} are given in the table. *See \cite[Fig.~25]{bloomberg19b}, reading the midpoint between the conservative and optimistic costs for electrolyzers in China in 2035 as 120~\$/kW.}
\label{tab:electrolysis_improvements}
\begin{tabular}{p{0.45\linewidth}p{0.45\linewidth}}
\hline\noalign{\smallskip}
Characteristic & Capital-cost (\$/kgH$_2$ per year) reduction factor\\
\noalign{\smallskip}\hline
MW-scale instead of tens of kW & 90--97\% (p.~23)\\
Prices in China vs.\ the West & 50--80\% (p.~23)\\
Continued learning by 2035 & 35\%* (p.~30)\\
Integration with hydrogen plant & 15\% (by avoiding grid connections, p.~64)\\
\noalign{\smallskip}\hline
\end{tabular}
\end{table}

Electrolysis can also be powered by renewables, albeit at a lower capacity factor. 
Many approaches are modeled in \cite[p.~52]{bloomberg19b},
and the findings are summarized in \cite[p.~67, Figs.~62 and 63]{bloomberg19b},
including that a combination of wind and solar can deliver low-cost power to an electrolyzer while still enabling a reasonable capacity factor. We model the cost of this system in \cite[Tab: Production, Column: China, optimized renewables 2030]{SH} and estimate
a cost of 1.52~\$/kgH$_2$. Fusion must produce power at $\le 32$~\$/\MWhe\ to compete with renewable production of hydrogen
in regions where wind and solar power are plentiful.

\subsubsection{Alternatives to Conventional Electrolysis}

We cannot find detailed costs for high-temperature electrolysis, although
the energy requirements for low- and high-temperature electrolysis are given in
\cite[p.~481]{badwal12}. If the capital costs are the same for both, we find that fusion LCOE would need to be 16--62~\$/\MWhe\ for high-temperature electrolysis to compete with traditional sources of hydrogen. This is only a modest improvement compared to low-temperature electrolysis
(requiring a fusion LCOE of 13--50~\$/\MWhe) because a significant fraction of the energy is still supplied as electricity. Furthermore, it is still an open research question whether a fusion power plant can deliver the necessary temperatures for high-temperature electrolysis.

Intermediate-temperature electrolysis (100--400$^\circ$C) 
would get a larger fraction of its energy from electricity than high-temperature electrolysis, and hence should be less of a win unless the heat can be supplied from waste heat from the fusion power plant. We found even fewer references for this, so we ignore intermediate-temperature electrolysis.

There are yet other ways to produce hydrogen from heat, including pyrolysis of natural gas, which uses heat to separate natural gas into hydrogen and carbon (not carbon dioxide)
\cite{serban03}, and thermo-chemical water splitting, which uses heat and chemical reactions to separate water into hydrogen and oxygen \cite{EERE}. Additionally, plasma catalysis uses electricity to produce hydrogen from methane \cite{carreon19}.  Detailed cost analyses of these options
are beyond the scope of this paper.

\subsubsection{Impact of a Carbon Price}

In most hydrogen markets that we examine below, green hydrogen does not compete with existing sources of hydrogen so much as with existing markets/processes that produce and consume hydrogen internally. For example, the production of ammonia involves both endothermic and exothermic processes, and is more efficient as a result of using heat from one to run the other. This produces a lower implied price for hydrogen than the market price. We examine the impact of a carbon price on specific markets in the next subsection.

Table~\ref{tab4} shows the cost impact of CCS on methane reformation and coal gasification, which emit significant amounts of CO$_2$ and suffer reduced efficiency 
if paired with CCS\@. 
Table~\ref{tab5} shows the impact of carbon pricing based on the numbers of
Table~\ref{tab4}. 
Without CCS, a price of 50~\$/tCO$_2$ would add 0.45 \$/kgH$_2$ to the cost of hydrogen based on methane reformation,
and more than 1~\$/kgH$_2$ based on coal.
At 50~\$/tCO$_2$, it is more economical for these plants to pay the carbon price rather than integrate CCS, whereas at 100~\$/tCO$_2$, the plants would save money by integrating CCS\@. A high carbon price can provide significant relief to green hydrogen, especially when it is competing with hydrogen produced from coal.

\begin{table*}[tb]
\caption{Impact of CCS on price of different hydrogen-production methods \cite{bloomberg20}; corresponding page numbers from \cite{bloomberg20} are given in the table.}
\label{tab4}
\begin{tabular}{lccc}
\hline\noalign{\smallskip}
Production method & Emissions  & Cost per 10~\$/tCO$_2$ of & Amortized cost of \\
 & (kgCO$_2$/kgH$_2$) & associated emissions (\$/kgH$_2$) & CCS (\$/kgH$_2$)\\
\noalign{\smallskip}\hline
Methane reformation & 8.9 (p.~4) & 0.089 & --\\
Methane reformation w/CCS & 0.89 (p.~11) & 0.0089 & 0.6 (p.~11)\\
Coal gasification & 20.2 (p.~4) & 0.202 & -- \\
Coal gasification w/CCS & 2.02 (p.~11) & 0.0202 & 1.1 (p.~11)\\
\noalign{\smallskip}\hline
\end{tabular}
\end{table*}

\begin{table}[tb]
\caption{Impact of various carbon prices on hydrogen prices, based on the information
in Table~\ref{tab4}.}
\label{tab5}
\begin{tabular}{ccc}
\hline\noalign{\smallskip}
Carbon price & Cost increase of  & Cost increase of\\
(\$/tCO$_2$) & methane reformation  & coal gasification \\
 & (\$/kgH$_2$) & (\$/kgH$_2$)\\
\noalign{\smallskip}\hline
20 & 0.18 (no CCS) & 0.40 (no CCS)\\
50 & 0.45 (no CCS) & 1.01 (no CCS)\\
100 & 0.69 (w/CCS) & 1.3 (w/CCS)\\
\noalign{\smallskip}\hline
\end{tabular}
\end{table}

\subsection{Uses of Hydrogen}

In this subsection, we examine some of the larger uses of hydrogen, not all of which are suitable
early markets for fusion. As noted above, many of these uses of hydrogen produce and consume 
hydrogen as part of a larger process, placing challenging cost constraints on green hydrogen 
production. We ignore new uses for hydrogen on the basis that they will be more price-sensitive 
than existing uses.

\subsubsection{Ammonia}

Hydrogen demand for ammonia production is predicted to be 38~MtH$_2$/year in 2030, i.e., 31~MtH$_2$/year in 2018 \cite[p.~99]{IEA19} and growing at 1.7\% per year from 2018--2030 \cite[p.~101]{IEA19}. While there are hopes for ammonia to play a major role as a fuel, today it is used almost exclusively in the chemical industry \cite[pp.~61 and 139]{IEA19},
and is mostly produced from natural gas or coal in an integrated and exothermic process from hydrogen and nitrogen. The energy released helps power the endothermic production of hydrogen by natural-gas reformation or coal gasification. The result is an implied price of 0.35--2.08~\$/kgH$_2$ when natural gas prices are in the range 2--12~\$/MMBTU \cite[p.~3]{bloomberg19c}, significantly less than the market price of green hydrogen. The highest price we find for ammonia production is in China, corresponding to 1.43~\$/kgH$_2$ and a
required fusion LCOE of 30~\$/\MWhe. Table~\ref{tab6} shows the five highest
costs from the top ten ammonia-producing countries.

\begin{table*}[tb]
\caption{Highest prices of ammonia (NH$_3$) from top producing countries \cite[Tab: Ammonia]{SH}. ``Equivalent fusion LCOE'' is the required
value for fusion to produce H$_2$ at the ``Required H$_2$ price,'' based on \cite[Tab: Production chart]{SH} and the assumptions of the ``Producing Hydrogen'' subsection.}
\label{tab6}
\begin{tabular}{lccccc}
\hline\noalign{\smallskip}
Interpolations & Quantity of NH$_3$ & Levelized cost of & Quantity of & Required H$_2$ & Equivalent fusion \\
 & (MtNH$_3$/yr) & NH$_3$ (\$/kgNH$_3$) & H$_2$ (MtH$_2$/yr) & cost (\$/kgH$_2$) & LCOE (\$/\MWhe)\\
\noalign{\smallskip}\hline
China  (natural gas) & 6 & 450 & 1.18 & 1.43 & 30\\
Ukraine & 5 & 430 & 0.94 & 1.33 & 27\\
China (coal) & 53 & 400 & 10.60 & 1.17 & 23\\
India & 13 & 370 & 2.60 & 1.01 & 20\\
Indonesia & 7 & 350 & 1.36 & 0.91 & 17\\
\noalign{\smallskip}\hline
\end{tabular}
\end{table*}

A carbon price provides significant breathing room for the 
price of hydrogen (summarized
in Table~\ref{tab7}). A 100~\$/tCO$_2$ price allows adding 0.9 and 1.5~\$/kgH$_2$ when competing with natural gas and coal, respectively \cite[p.~7]{bloomberg19c}.  
We estimate that a
50~\$/tCO$_2$ price would raise the competitive price for the 
bulk of the market by about 0.5--1.17~\$/kgH$_2$. Unless the carbon pricing is carefully designed, it
will not capture CO$_2$ emissions from the use of
ammonia-based fertilizers, reducing its
effect \cite[p.~7]{bloomberg19c}.

\begin{table}[tb]
\caption{Ammonia market summary. ``Fusion LCOE'' is the required value for fusion to produce H$_2$ at the ``Hydrogen price,'' based on \cite[Tab: Production chart]{SH} and the assumptions of the ``Producing Hydrogen'' subsection.}
\label{tab7}
\begin{tabular}{cccc}
\hline\noalign{\smallskip}
Carbon & Fusion & Hydrogen & Market \\
price &  LCOE & price & size\\
(\$/tCO$_2$) & (\$/\MWhe) & (\$/kgH$_2$)  & (MtH$_2$/year) \\
\noalign{\smallskip}\hline
0 & 12 & 0.67 & 27\\
20 & 17 & 0.87 & 27\\
50 & 23 & 1.17 & 27\\
100 & 35 & 1.67 & 27\\
\noalign{\smallskip}\hline
\end{tabular}
\end{table}

\subsubsection{Methanol}

Hydrogen demand for methanol production in 2030 is predicted to be 18~MtH$_2$/year, i.e., 12~MtH$_2$/year in 2018 \cite[p.~99]{IEA19} and
growing at 3.6\% per year from 2018--2030 
\cite[p.~101]{IEA19}. Methanol (CH$_3$OH) is widely used, e.g., in plastics and for blending into liquid fuels. However, the market is inaccessible to green hydrogen because the majority of methanol is produced in areas with very low-cost natural gas and because of the cost of acquiring CO$_2$ feedstock that supplies the carbon in the methanol.

For example, producing methanol from green hydrogen that costs 1~\$/kgH$_2$ and carbon dioxide that costs 95~\$/tCO$_2$ cannot compete with methanol made from the most expensive natural gas or coal \cite[p.~7]{bloomberg19d}. With a carbon price, the methanol facility can be paid for the CO$_2$ that it uses, instead of having to pay for it. At 100~\$/tCO$_2$, the maximum carbon price that we consider, and with hydrogen at 2~\$/kgH$_2$, a methanol plant can compete with the most expensive natural gas, but hydrogen under 1~\$/kgH$_2$ is still required to compete with lower-cost natural gas. A 100~\$/tCO$_2$ carbon price makes green methanol more competitive with methanol made from coal: it only requires hydrogen between approximately 2--3~\$/kgH$_2$ to compete with low-cost and expensive coal, respectively \cite[p.~9]{bloomberg19d}.
Perhaps with a 100~\$/tCO$_2$ carbon price and low-cost hydrogen, this can be made to work, but the market seems difficult enough that we ignore it.

\subsubsection{Oil refining}

Refineries currently use 38~MtH$_2$/year \cite[p.~9]{IEA19}, which is expected to grow 7\% by 2030 \cite[p.~95]{IEA19}. About 20\% of this is supplied by commercial sources \cite[p.~95]{IEA19} at the market prices described in the ``Producing Hydrogen'' subsection above. It is possible that there are some refineries that pay significantly more for hydrogen, but we were unable to find data on this.
The rest of the hydrogen consumed is produced onsite in already paid-for facilities or as a zero-cost byproduct of oil refining, both of which are difficult to undercut. We assume that there will not be significant construction of new oil refineries.  A summary of the oil-refining
market is shown in Table~\ref{tab8}.

\begin{table}[tb]
\caption{Oil refining market summary. ``Fusion LCOE'' is the required value for fusion to produce H$_2$ at the ``Hydrogen price,'' based on \cite[Tab: Production chart]{SH} and the assumptions of the ``Producing Hydrogen'' subsection.}
\label{tab8}
\begin{tabular}{cccc}
\hline\noalign{\smallskip}
Carbon & Fusion & Hydrogen & Market \\
price & LCOE & price & size\\
(\$/tCO$_2$) & (\$/\MWhe)& (\$/kgH$_2$) & (MtH$_2$/year) \\
\noalign{\smallskip}\hline
0 & 20 & 1.00 & 8\\
20 & 24 & 1.18 & 8\\
50 & 30 & 1.45 & 8\\
100 & 36 & 1.69 & 8\\
\noalign{\smallskip}\hline
\end{tabular}
\end{table}

\subsubsection{Iron and Steel Refining}

Blast furnaces (BF) run on coal, do not need to be supplied with hydrogen, and produce 90\% of the world’s iron \cite[p.~108]{IEA19}. A less-common method of iron production is direct reduction (DR), which runs on natural gas but can run on hydrogen instead. Coal and gas both have wide price ranges but are primarily used in places where they have low cost, so the low end of the price range is the relevant one.

Without carbon pricing, green hydrogen can compete with DR powered by the lowest-cost natural gas (2~\$/MMBTU) at about 0.7~\$/kgH$_2$, opening up a 4-MtH$_2$/year market. Hydrogen must drop to 0.5~\$/kgH$_2$ to compete with a blast furnace run on even moderately low-cost coal \cite[p.~8]{bloomberg19e}. Because this price is very low and the blast furnaces would need to be replaced with DR equipment to run on hydrogen, we ignored this larger market in scenarios without carbon pricing.

Carbon pricing significantly reduces the price pressure on hydrogen. A 100 and 50~\$/tCO$_2$ price makes 2 and 1~\$/kgH$_2$, respectively, competitive with all BF steel and most DR steel, regardless of the fuel price. If all BF facilities converted to DR, then this would grow the market to 40~MtH$_2$/year. In the market summary (Table~\ref{tab9}), we assume that all facilities would convert to DR and use green hydrogen in scenarios with a carbon price. A carbon price of 20~\$/tCO$_2$ would require 0.7~\$/kgH$_2$ to compete with the lowest-cost gas \cite[p.~8]{bloomberg19e}.

\begin{table}[tb]
\caption{Iron and steel refining market summary. ``Fusion LCOE'' is the required value for fusion to produce H$_2$ at the ``Hydrogen price,'' based on \cite[Tab: Production chart]{SH} and the assumptions of the ``Producing Hydrogen'' subsection.}
\label{tab9}
\begin{tabular}{cccc}
\hline\noalign{\smallskip}
Carbon & Fusion & Hydrogen & Market \\
price & LCOE & price & size\\
(\$/tCO$_2$) & (\$/\MWhe) & (\$/kgH$_2$) & (MtH$_2$/year) \\
\noalign{\smallskip}\hline
0 & 8 & 0.5 & 4\\
20 & 13 & 0.7 & 4\\
50 & 20 & 1 & 40\\
100 & 43 & 2 & $\sim 40$\\
\noalign{\smallskip}\hline
\end{tabular}
\end{table}

\subsubsection{Heating Buildings}

Using pure hydrogen to heat buildings is challenging because it requires upgrades to all the equipment in the gas network and coordination between many parties \cite[p.~144]{IEA19}.
However, various studies have found that hydrogen can be mixed with natural gas up to at least 30\% (by volume) without causing problems
\cite[p.~147]{IEA19}, and markets appear able to support high prices. Table~\ref{tab11} shows the size of the hydrogen market for heating buildings, assuming a 15\% hydrogen mixture and a
price range for hydrogen to be competitive. Notably, the market starts to open up at 78~\$/\MWhe\ in Japan and 66~\$/\MWhe\ in Europe. We do not have data on the size of the market in these countries at each price point, but they should be large enough to support several early fusion plants \cite[p.~149]{IEA19}.
A carbon price adds 0.09~\$/kgH$_2$ \cite[Tab: Heating]{SE} to the allowable hydrogen prices for every 10~\$/tCO$_2$
(Table~\ref{tab11}). If environmental or economic pressures are large enough, these markets might be able to double in size if they use 30\% hydrogen instead of 15\%.

\begin{table*}[tb]
\caption{Hydrogen market for heating buildings \cite[Tab: Heating]{SH}.}
\label{tab10}
\begin{tabular}{p{0.07\linewidth}ccccccc}
\hline\noalign{\smallskip}
Region & Price, high & Price, high & Price, low & Price, low & Natural gas & H$_2$ demand at & Electricity\\
 & (\$/kgH$_2$) & (\$/\MWhe) & (\$/kgH$_2$) & (\$/\MWhe) & demand for heating & 15\% of gas & consumption to\\
 & & & & & (Mtoe/year) & (MtH$_2$/yr) & produce (GW)\\
\noalign{\smallskip}\hline
Japan & 3.5 & 78 & 2 & 43 & 14 & 0.73 & 4.19\\
Western Europe & 3 & 66 & 2 & 43 & 80 & 4.19 & 23.92\\
Korea & 1.9 & 40 & 0.9 & 17 & 11 & 0.58 & 3.29\\
Russia & 1.8 & 38 & 1.5 & 31 & 43 & 2.25 & 12.86\\
United States & 1.5 & 31 & 1.2 & 24 & 147 & 7.70 & 43.95\\
China & 1.4 & 29 & 1.2 & 24 & 51 & 2.67 & 15.25\\
Canada & 1.2 & 24 & 0.8 & 15 & 21 & 1.10 & 6.28\\
\noalign{\smallskip}\hline
\end{tabular}
\end{table*}

\begin{table}[tb]
\caption{Heating buildings market summary. ``Fusion LCOE'' is the required value for fusion to produce H$_2$ at the ``Hydrogen price,'' based on \cite[Tab: Production chart]{SH} and the assumptions of the ``Producing Hydrogen'' subsection.}
\label{tab11}
\begin{tabular}{cccc}
\hline\noalign{\smallskip}
Carbon & Fusion & Hydrogen & Market \\
price & LCOE & price & size\\
(\$/tCO$_2$) & (\$/\MWhe) & (\$/kgH$_2$) & (MtH$_2$/year) \\
\noalign{\smallskip}\hline
0 & 43 & 2 & 5\\
0 & 17 & 0.9 & 19\\
20 & 20 & 1.0 & 19\\
50 & 24& 1.2 & 19\\
100 & 31 & 1.5 & 19\\
\noalign{\smallskip}\hline
\end{tabular}
\end{table}

\subsubsection{High-Grade Heat for Other Purposes}

There is a large potential market, 370~Mtoe/year or 500 GW, for using green hydrogen to produce high-grade process heat, primarily for producing cement and various chemicals \cite[p.~116]{IEA19}. However, hydrogen is not competitive with other sources of heat even with a carbon price of 100~\$/tCO$_2$
\cite[p.~118]{IEA19}, and the market is hard to address because of significant variation in equipment even within an industry
\cite[p.~2]{bloomberg19a}. Accordingly, we conclude that there is no early fusion market for this.

\section{Economic Boosts}
There are several promising options for improving the economics of a fusion plant, including cogeneration using the waste heat and by reducing the capital cost.

\subsection{Cogeneration}

Revenue extracted from waste heat after generating electricity,
i.e., cogeneration, can reduce the effective LCOE and make a plant more competitive. 
This subsection explores the feasibility and profitability of cogenerating various products from fusion power plants.

\subsubsection{Desalination}

The increased need for desalination facilities in the future will create additional opportunities for fusion energy. Assuming an LCOE of 50~\$/\MWhe\ for the fusion power plant, integrating a fusion and thermal desalination plant could lower the effective fusion LCOE by approximately 30\%, depending on the specific desalination method.

Due to over-exploitation of resources and effects of climate change, fresh-water supplies around the world are declining. At the same time, population and economic growth will increase the demand for fresh water, with much of this growth in areas that already lack access to clean water. Global demand for all water uses, which is currently around 4600 km$^3$/year, is predicted to increase by 20--30\% by 2050 to 5500--6000~km$^3$/year \cite{boretti19}. In addition, it is also projected that by 2050, 57\% of the global population will live in areas that suffer water scarcity at least one month each year \cite{boretti19}. While water efficiency measures and further regulations could help, additional sources of clean water may be needed.

Desalination, which uses energy to remove salt from water to produce water suitable for drinking or agriculture, can be a solution to worsening water shortages. The first large-scale desalination plants were built in the 1960s, and there are about 20,000 facilities in use today \cite{robbins19}. Because of the projected increase in demand for water, it is likely that desalination will become a larger portion of the water market by the time fusion energy is commercially viable. If the predicted 20\% demand increase by 2050 is entirely supplied by desalination, this would create a need for
approximately 700--1000 GW (depending on the type of desalination used) \cite[Tab: Desalination Analysis]{SE} in order to power the new desalination plants (energy needs are from \cite[p.~40]{world_bank_desal}).

There are three main types of desalination technology. These include two thermal processes, Multistage Flash Distillation (MSF) and Multi-effect Distillation (MED), and one membrane process, Seawater Reverse Osmosis (SWRO)\@. For MSF and MED desalination plants, heat and electricity are 52\% and 14\% of the O\&M costs, respectively. For SWRO desalination plants, electricity is 41\% of O\&M costs, with no heat required 
\cite[p.~27]{world_bank_desal}. MSF technology can produce water at scale at lower cost than MED technology, which is more common for smaller thermal desalination plants \cite[p.~21]{world_bank_desal}. SWRO also can offer significant economies of scale at smaller sizes, but those taper off above plant capacities of 100 MLD
(million liters per day). However, with no heat requirements, SWRO alone will likely not be the best fit for fusion. Some desalination plants are also built as hybrid plants, in order to incorporate a combination of either MSF or MED with SWRO, thus allowing for the plant to take advantage of the lowest-cost available thermal or electrical energy, while still allowing 24-hour water production.

We estimate the savings from powering a desalination plant using waste heat from a fusion power plant. Both thermal desalination methods require low-grade heat: around $100^\circ$C and  $70^\circ$C for MSF and MED, respectively \cite[p.~12]{IRENA12}. Fusion can take advantage of these requirements by supplying waste heat at little-to-no cost. The cost components of an MSF and MED plant that can deliver water at 1.07 \$/m$^3$ and 0.83~\$/m$^3$, respectively, are described
in \cite[p.~47]{world_bank_desal}, assuming an LCOE of 50~\$/\MWhe. We assume that the fusion power plant, which is sized to supply all its waste heat to meet the desalination requirements, provides heat and electricity to the desalination plant, and that each \MWht\ used by the desalination plant reduces the electrical output of the fusion power plant by 0.15~\MWhe\ \cite[p.~50]{qvist20a}. We then estimate the revenue that the fusion power plant receives from desalinated water as the difference between the cost of running the desalination plant with grid connections and the cost of running it with heat and electricity from the fusion power plant. Incorporating these factors, we find that cogeneration using MSF and MED lowers the effective fusion LCOE 27\% (from 50 to 37~\$/\MWhe) and 35\% (to 32~\$/\MWhe), respectively \cite[Tab: Desalination Analysis]{SE}. 

While this analysis may make MED appear to be the best desalination type to combine with fusion, in reality it depends on the size of the desalination facility. Although MSF has slightly higher capital costs and energy requirements than MED, it is easier to operate and benefits more from economies of scale at higher capacities, such as 500~MLD or greater \cite[p.~45]{world_bank_desal}. Although this would be considered a mega-size desalination project, a plant would likely need to be in this size range in order to economically pair with fusion. For example, based on average energy requirements for MSF, a 225-MW$_\mathrm{e}$ fusion plant would be able to deliver all its energy to a 500-MLD desalination plant \cite[Tab: Desalination Analysis]{SE}. Using fusion energy to power desalination could provide a unique opportunity to open additional markets for fusion, reduce the effective costs of a fusion plant, and produce desalinated water in a clean and sustainable way.

\subsubsection{Direct Air Capture}

Powering a direct-air-capture (DAC) plant with waste heat from a fusion power plant could reduce the effective fusion LCOE\@. Assuming contemporary costs of a DAC plant and an
LCOE of 50~\$/\MWhe\ for the fusion power plant, integrating the two plants could lower the effective fusion LCOE by about 35\%.

According to the IPCC, any chance to keep the global average temperature increase from pre-industrial levels under $1.5^\circ$C will require some form of CO$_2$ removal \cite{IPCC18}. One solution for removing CO$_2$ is DAC, which pulls CO$_2$ directly from the atmosphere at a much lower concentration than point-source carbon capture. DAC technology is relatively new, energy-intensive, and expensive.  However, pilot plants have been built, and new technologies with increased scale will reduce costs. It is also likely that by the time fusion energy becomes commercially viable, the demand for DAC may be much higher \cite{hanna21}. A conservative estimate for negative emissions eventually needed to reach global climate targets is about 10~GtCO$_2$ per year \cite{peters17}. If this amount were to come entirely from the solid-sorbent DAC method (see below), it would require roughly 1300 GW of heat and 270 GW of electricity. This increase in energy demand could become a significant market for fusion energy \cite[Tab: DAC Analysis]{SE}.

Currently, there are two main types of DAC technology: a high-temperature, liquid-solvent method, and a low-temperature, solid-sorbent method. Both require mostly heat, as well as some electricity. The liquid-solvent and solid-sorbent methods require heat at approximately $900^\circ$C and $100^\circ$C, respectively \cite[p.~962]{fasihi19}. Because the upper temperature limit for early fusion power plants is not yet known, we focus on the solid-sorbent method in our analysis.

Consider the example of a solid-sorbent DAC plant described in \cite[p.~965]{fasihi19}. The plant captures 360,000 tons of CO$_2$ per year at 187~\$/tCO$_2$, assuming an LCOE of 50~\$/\MWhe\ and a levelized cost of heat (LCOH) of 20~\$/\MWht. We assume that the fusion power plant, which is sized to supply all its waste heat to meet the DAC requirements, provides heat and electricity to the DAC plant, and that each \MWht\ used by the DAC plant reduces the electrical output of the fusion power plant by 0.15~\MWhe\ \cite[p.~50]{qvist20a}. We then estimate the revenue that the fusion power plant receives from captured CO$_2$ as the difference between the cost of running the DAC plant with grid electricity and the cost of running it from heat and electricity from the fusion power plant.
(In this analysis, the price of DAC when paying for energy was more than our 100-\$/tCO$_2$ threshold for a carbon price. In the long run, we expect prices to be lower, but we used the best data that we had for this analysis.)
Incorporating all these factors, we found that even with a reduction in electricity sold, the revenue from the DAC plant lowers the effective fusion LCOE from 50 to 32~\$/\MWhe, i.e., a 35\% decrease \cite[Tab: DAC Analysis]{SE}. Although this analysis has uncertainty, it suggests that using a fusion power plant to power a DAC plant could reduce the effective fusion LCOE, assuming that there is a market for captured CO$_2$.

Lastly, when compared to desalination, DAC as an economic boost for fusion has somewhat lower near-term potential. This is because both analyses depend on the assumption that the plants take in revenue equal to the cost of capturing carbon or producing desalinated water. Athough the situation could easily change over the next few decades, there is presently very little market demand or financial incentive to pay for CO$_2$, whereas there is already a well-established and increasing demand for fresh water in many areas.

\subsubsection{District Heating}

Powering a district-heating system with fusion appears to be a promising long-term market, but in the short term it requires a large, pre-existing heat network, which seems to be a rare opportunity.

District heating provides heat to buildings from a shared heat source, which produces steam or hot water that is piped to each building and heat-exchanged with the building’s heating system. This can be an economical alternative to
on-site combustion of fossil fuels or heating via electricity (an expensive form of energy), especially when using heat sources that do not scale down to the power required by a single building, e.g., fission or fusion power plants.
In the long term, it almost certainly makes sense for fusion to power district-heating networks. The world uses 17.5~PWh$_\mathrm{t}$ of low-grade ($<200^\circ$C) heat annually \cite[p.~48]{IAEA17}. Individual cities in northern Europe could each use multiple GW$_\mathrm{t}$ of district heating \cite[p.~58]{qvist20a}. When provided by a fission or fusion power plant, each \MWht\ extracted only reduces electricity production by 0.15~MWhe \cite[p.~50]{qvist20a}. If fusion power plants can be sited close to population centers, the pipes between the heat source and the heat market could be shortened, saving hundreds of millions to billions of dollars \cite[p.~55]{qvist20a}.

However, it is unclear that district heating makes sense as an early market for fusion. Existing heat networks are small: 17,000 networks in the UK together carry 1.4~GW in total \cite[p.~48]{qvist20a}, and where they are heated by fission power plants, they use an average of 5\% of the plant’s heat
\cite[p.~48]{IAEA17}. We find it unlikely that using 5\% of the waste heat of a fusion power plant for heating will change the economics of the plant. However, a coal power plant with a significant district-heating load on its waste heat would be an excellent candidate for retrofitting with a fusion power plant, capturing extra revenue from heating as well as cost savings from reusing part of the coal power plant, as detailed in the ``Retrofitting Coal Plants'' section below.

To be suitable for district heating, a fusion power plant needs to be able to supply hot water or steam at $<200^\circ$C. Modifying existing fission power plants to do this is relatively straightforward \cite[p.~48]{IAEA17}, and thus should probably not be a problem for fusion plants either. Given that this market remains small for now, we recommend that fusion developers focus on this only if it proves necessary to make the economics work and/or find a district-heating opportunity that uses a large fraction of the waste heat.

\subsection{Lowering Capital Costs}

In order to fully compete with NGCC power plants, fusion companies should look to minimize the overall capital that must be risked to build a fusion power plant.
Fuel is 80\% of the LCOE of an NGCC power plant, which makes the construction of a new plant relatively low financial risk. By contrast, capital costs are 80\% of the LCOE from fission \cite[p.~34]{MIT18} and will likely be similar for fusion. Even if two power plants, e.g., one fusion and one NGCC, have the same LCOE, the fusion power plant will be a much larger financial risk because more of the costs committed during construction will be lost if the plant is never completed or becomes unprofitable to operate.
This fact is partly hidden by the metrics used by the industry.

\subsubsection{Overnight vs.\ Total Construction Cost}

When considering the cost to construct a plant, it is important to use the correct metrics. 
Often, power-plant developers quote the overnight construction
(or capital) cost (OCC), which is the cost in dollars per watt (\$/W) required to build a plant if it could be constructed instantaneously, i.e., ``overnight.'' The first problem with OCC is that it ignores construction time and the cost of capital,
which can be substantial. For example \cite[p.~17]{CATF18}, a fission power plant that takes 10 years to build with an 8\% interest rate (a high-risk project) will cost 70\% more than an NGCC power plant that takes two years to build and has a 6\% interest rate, even though they have the same OCC \cite[Tab: OCC/TCC]{SE}. The second problem is that OCC does not measure financial and bankruptcy risks.  The latter are better measured by the total construction (or capital) cost (TCC), which drives decisions about whether plants get built. TCC, measured in dollars rather than \$/W, is the sum of all 
direct and indirect (including interest) costs until the plant is producing revenue. Fusion developers should avoid the pitfall of designing a very large plant to achieve low \$/W\@.  
Determining a range of recommended fusion TCCs is beyond the scope of this report, but
we recommend adopting methods to minimize construction time and TCC, e.g., see \cite[Sec.~2.5]{MIT18}.

\subsubsection{Retrofitting Coal Plants}

Another potential way to reduce TCC is to reuse equipment in a coal power plant. A fusion core that can deliver steam at 500--600$^\circ$C can save up to 30\% on TCC by retrofitting an existing coal plant. Because fusion LCOE is likely to be dominated by capital costs, retrofitting a coal plant could also significantly reduce LCOE\@.

A recent study \cite{qvist21} examined retrofitting an existing coal power plant to generate clean power through various combinations of solar, wind, enhanced geothermal systems, biomass, CCS, or small modular fission reactors. The study found that fission reactors would be the most viable retrofit option for coal power plants while generating a similar amount of electricity. 
However, there are some caveats to this finding, as it would not be useful to retrofit coal plants of any age or size. The most applicable coal plants are ones with greater than 50-MW capacity and either recently built or modernized, meaning most of the valuable equipment on the site is less than 20 years old \cite[p.~12]{qvist21}. Although likely reduced by the time fusion is commercially viable, there are currently about 1,100 coal plants worldwide with a combined capacity of roughly 1.1~TW that fall into this category
\cite{WWI19}. It is worth noting that about 800 of these coal plants are in China. In addition, some have capacities in the range of multiple GW, which could allow for multiple fusion cores to be built on the same site.

To maximize the cost savings of installing a fusion energy system at a coal site, all applicable equipment should be reused. Ordered from easiest to most difficult, this includes the site location, grid connection, electrical equipment, external heat interfaces, turbine and generator, and steam-cycle cooling equipment. In order to reuse the full steam cycle, the thermal heat output of the fusion plant would need to generally match, or be a multiple of, the thermal output of the coal unit \cite[p.~17]{qvist21}. It should be noted that even with reusing everything up to the full steam cycle, only 40\% of the coal-plant costs will be recovered, as the other 60\% is dominated
by the coal, ash, and flue-gas systems that are not needed \cite[p.~7]{qvist21}. 

By comparing the savings anticipated when repowering a 460-MW$_\mathrm{t}$ coal power plant with an advanced fission reactor \cite[p.~27]{qvist21} against cost estimates from an
update \cite{woodruff20} to a study \cite[pp.~25--26]{woodruff17} for a 150-MW$_\mathrm{e}$
fusion power plant
(corresponding to 460~MW$_\mathrm{t}$ assuming 33\% conversion efficiency), we find that a fusion retrofit might save 
22--40\% of capital costs across four fusion plant designs 
\cite[Tab: Repowering coal plants]{SE}.  Furthermore,
significant job retention may be possible compared to
abandoning coal plants as a stranded asset \cite[p.~33]{qvist21}.


While there are uncertainties needing more careful examination, it appears that TCC can be meaningfully reduced by retrofitting a coal power plant with fusion.

\section{Summary}

In this paper, we examined cost requirements for
fusion-generated electricity, process heat, and hydrogen production 
based on today's market prices but
with various adjustments relating to possible
scenarios in 2035, such as
``business-as-usual,'' high renewables penetration, and carbon 
pricing up to 100~\$/tCO$_2$.
We took a relatively unforgiving view (in light of anticipated 
continued cost reductions of the competition such as renewables) in 
order to be conservative,
resulting in generally aggressive cost targets that fusion must hit 
to break into its first markets.

Key findings are that fusion developers should consider focusing initially on high-priced
global electricity markets and including integrated thermal storage in order
to maximize revenue and compete in markets with high renewables penetration. Process heat
and hydrogen production will be tough early markets for fusion, but may open
up to fusion as markets evolve and if fusion's LCOE falls below 50~\$/\MWhe. Fusion 
plants could potentially increase revenue
via cogeneration (e.g., desalination, direct air capture, or
district heating) and lower capital costs (e.g., by minimizing construction times and interest or by retrofitting coal plants).
Findings are summarized in greater detail in the 
``Key Findings" section of the paper.

Finally, although this paper focuses on commercial energy-related
markets,
fusion may also have a range of space applications, e.g., see \cite{cassibry15,wurden16},
with different techno-economic requirements.  Assessing
space applications as a potential early market for fusion was beyond the scope of this paper.

\begin{acknowledgements}
We are grateful for the insights and input provided by many people, especially Bob Mumgaard, Brandon Sorbom, Shiaoching Tse, and Ally Yost (Commonwealth Fusion Systems), Joe Chaisson (Clean Air Task Force, Energy Options Network), Eric Ingersoll (Lucid Catalyst), Armond Cohen (Clean Air Task Force), and Richard Pearson (Kyoto Fusioneering).
We thank Jennifer Steinhilber (Booz Allen Hamilton) for assistance with several of the figures.
Reference herein to any specific non-federal person or commercial entity, product, process, or service by trade name, trademark, manufacturer, or otherwise, does not necessarily constitute or imply its endorsement, recommendation, or favoring by the U.S. Government or any agency thereof or its contractors or subcontractors.
\end{acknowledgements}

%
%


%
%

\end{document}